\documentclass[sigconf]{acmart}

\usepackage{color}
\usepackage{algorithm}
\usepackage{algorithmic}
\usepackage{multirow}
\usepackage{enumerate}
\usepackage{url}
\usepackage{graphicx}

\usepackage{subfloat}
\usepackage{mwe}
\usepackage{makecell}

\usepackage{subfig}
\usepackage{float}
\usepackage{array}
\usepackage{booktabs}
\usepackage{pslatex}
\usepackage{amssymb}

\usepackage{pifont}
\newcommand{\cmark}{\ding{51}}%
\newcommand{\xmark}{\ding{55}}%

\usepackage[T1]{fontenc}

\begin{document}

\title{Recognising Cardiac Abnormalities in Wearable Device\\ Photoplethysmography (PPG) with Deep Learning}

\newcommand{\todo}[1]{\textcolor{red}{#1}}
\newcommand{\argmax}{\operatornamewithlimits{arg\,max}}
\renewcommand{\labelitemi}{$\bullet$}

	\author{Stewart Whiting, Samuel Moreland, Jason Costello, Glen Colopy,  Christopher McCann}
\affiliation{%
  \institution{snap40, 24 Forth Street, Edinburgh, Scotland, UK.}
}
\email{ [stewart,sam,jason,glen,christopher]@snap40.com}

\begin{abstract}

Cardiac abnormalities affecting heart rate and rhythm are commonly observed in both healthy and acutely unwell people. 
Although many of these are benign, they can sometimes indicate a serious health risk. 
ECG monitors are typically used to detect these events in electrical heart activity, however they are impractical for continuous long-term use. In contrast, current-generation wearables with optical photoplethysmography (PPG) have gained popularity with their low-cost, lack of wires and tiny size.
Many cardiac abnormalities such as ectopic beats and AF can manifest as both obvious and subtle anomalies in a PPG waveform as they disrupt blood flow.
We propose an automatic method for recognising these anomalies in PPG signal alone, without the need for ECG.
We train an LSTM deep neural network on 400,000 clean PPG samples to learn typical PPG morphology and rhythm, and flag PPG signal diverging from this as cardiac abnormalities.
We compare the cardiac abnormalities our approach recognises with the ectopic beats recorded by a bedside ECG monitor for 29 patients over 47.6 hours of gold standard observations.
Our proposed cardiac abnormality recognition approach recognises 60\%+ of ECG-detected PVCs in PPG signal, with a false positive rate of 23\% -- demonstrating the compelling power and value of this novel approach. 
Finally we examine how cardiac abnormalities manifest in PPG signal for in- and out-of-hospital patient populations using a wearable device during standard care.

\end{abstract}

\acmDOI{10.475/123_4}

\acmISBN{123-4567-24-567/08/06}

\acmConference[Workshop on Machine Learning for Medicine \& \\ Healthcare, KDD 2018]{ACM SIGKDD conference}{London}{UK}
\acmYear{2018}
\copyrightyear{2018}

\acmArticle{4}
\acmPrice{15.00}

\settopmatter{printacmref=false} 
\renewcommand\footnotetextcopyrightpermission[1]{}

\maketitle

\section{Introduction}
\label{sec:introduction}

Many cardiac abnormalities affecting heart rate and rhythm are observed in both healthy and acutely unwell populations.
These often present through arrhythmias, where the heart beat is either persistently or occasionally irregular, too fast or too slow.
While there are many different types of arrhythmias, among the most common are tachycardia (i.e., >120 beats/min), bradycardia (i.e., <45beats/min), atrial fibrillation (AF) and flutter (i.e., disordered and fluctuating heart beats) and ectopic beats such as premature ventricular and atrial contractions (i.e., PVCs/PACs).
%
%
%
%

Many arrhythmias are asymptomatic or benign, and can occur in otherwise healthy individuals seemingly randomly. 
However, at increasing frequency for patients in high risk groups, they may be a precursor to, or part of an acute condition. AF has been shown to predispose a patient to stroke or heart failure. PVCs have been shown to manifest with cardiomyopathy and myocardial infarction.

An electrocardiogram (i.e., \textit{ECG}, or \textit{EKG}) measuring electrical activity in the heart over time is the main diagnostic approach for investigating cardiac abnormalities. 
An ECG examination requires a patient to have 2, 6 or 12 leads connected to electrodes correctly attached to skin across their chest and limbs. 
%
%
%
%
%
%
Bedside and portable telemetry Holter ECG monitors are the clinical gold standard for accurate cardiac diagnosis. However, 
they are impractical or uncomfortable for long-term continuous use in ambulatory patients, so ECG investigations typically only take place over relatively short periods of time for patients who have presented other symptoms which warrant the investigation. As a result, many rarer, asymptomatic or early-onset cardiac conditions can be missed.

In contrast, conveniently small, unobtrusive and inexpensive wearable devices such as smart watches and fitness trackers which include a photoplethysmography (PPG) sensor to monitor the user's pulse have become extremely popular.
PPG is an optical sensing technique which transmits specific light wavelengths into well-perfused skin tissue, and measures the amount of light reflected back - thereby measuring the changing volume of blood in the tissue over time following each pulse wave ejected from the heart. 
A typical real wearable device pulsatile PPG waveform is shown in Figure 1(a).


\begin{figure}
   \centering
   \subfloat[][]{\includegraphics[width=.22\textwidth]{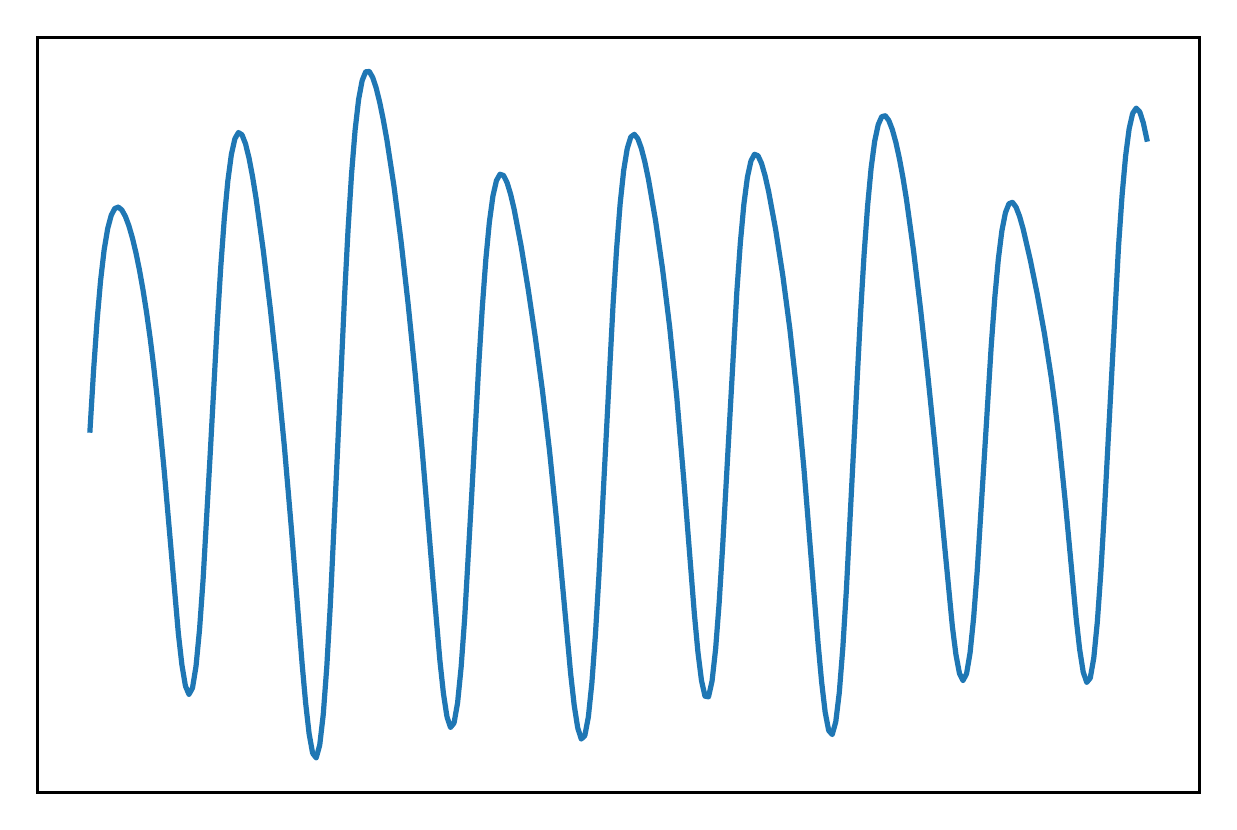}}\quad
    \subfloat[][]{\includegraphics[width=.22\textwidth]{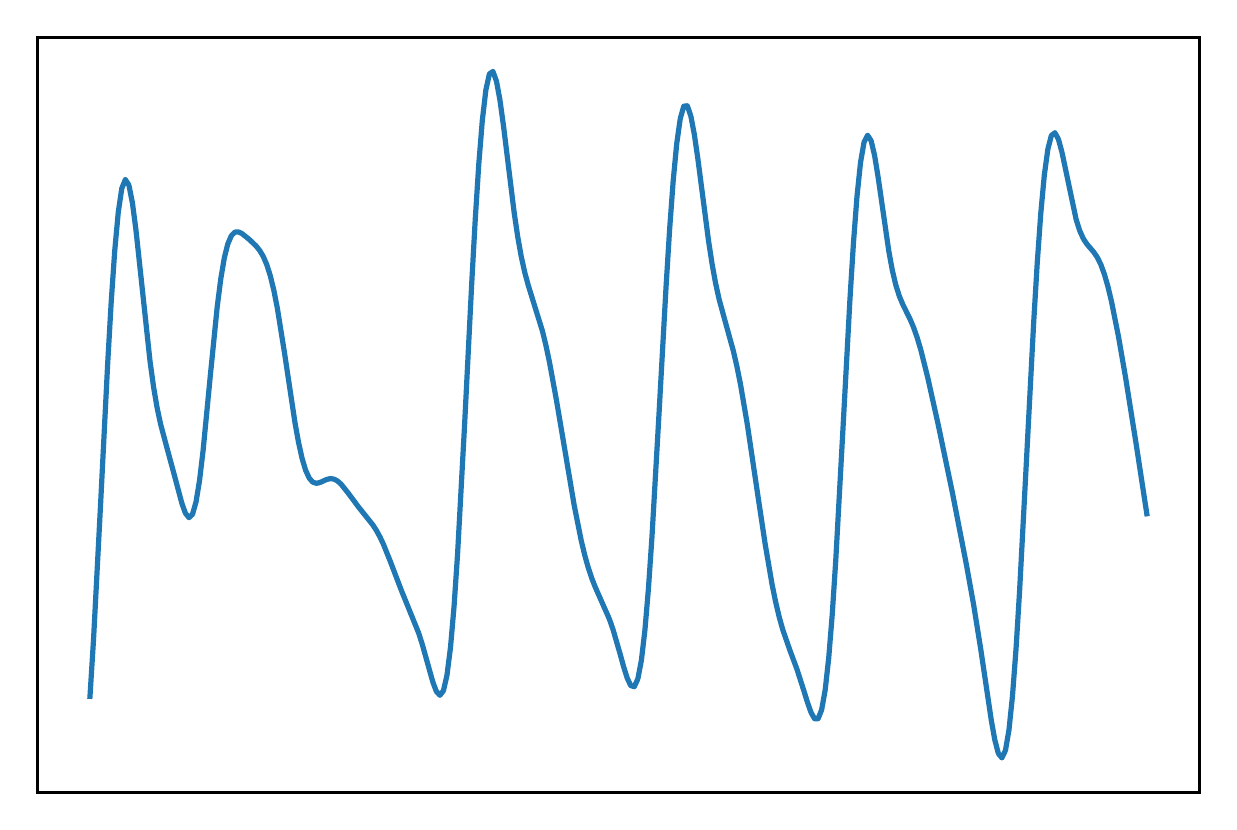}}\\
      \vspace{-0.2cm} 
       \subfloat[][]{\includegraphics[width=.22\textwidth]{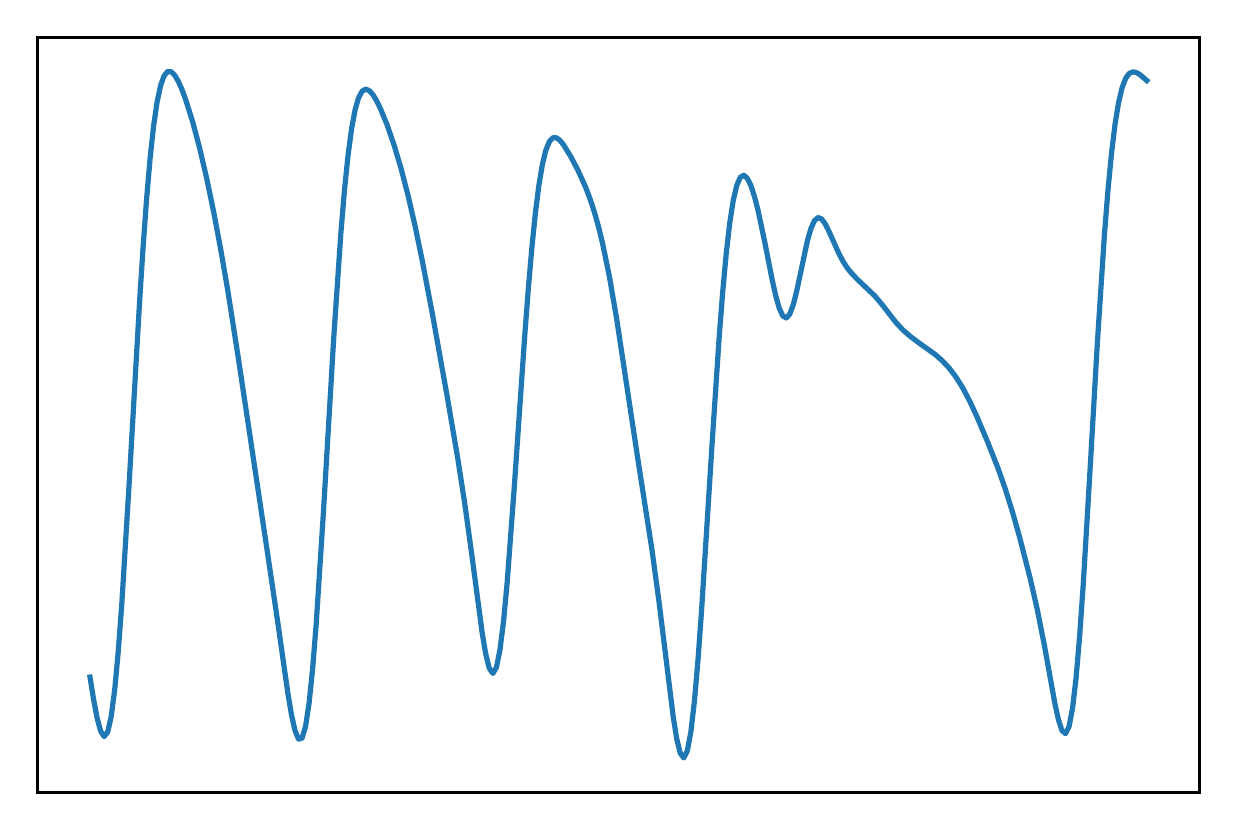}}\quad
    \subfloat[][]{\includegraphics[width=.22\textwidth]{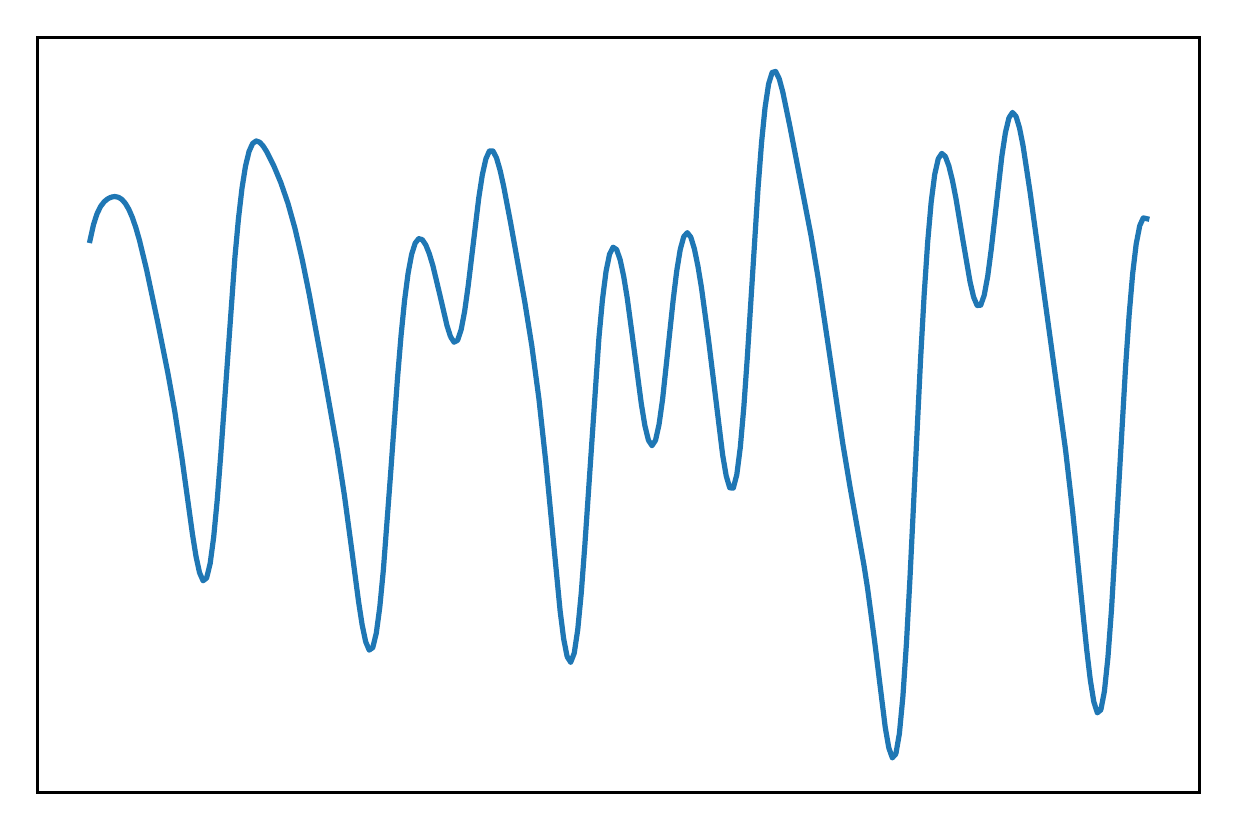}}
  \vspace{-0.2cm} 
   \caption{6-second PPG signal samples. (a) is a regular rhythmic PPG signal, while (b), (c) and (d) contain cardiac anomalies.}
   \label{fig:anomalousppgexamples}
\vspace{-0.4cm}

\end{figure}

Under normal conditions, the heart atria and ventricles contract sequentially to pump blood into the arterial system. Under abnormal conditions there can be a mistiming of contractions (e.g., a PVC), which can cause faster rhythm and reduced cardiac output. Similarly, reduced or no atrial contraction with erratic ventricular contraction (e.g., AF) can lead to random heart rate and cardiac output.

As PPG only measures the output of the heart into the circulatory system, it cannot fully characterise the underlying heart activity which preceded it as with the fidelity of an ECG.
%
%
%
However, since many cardiac abnormalities affect the heart's pulse wave output, they can disrupt in various ways -- albeit sometimes subtly -- blood flow and thus `glitch' subsequent PPG waveform rhythm and morphology. Real examples of brief cardiac anomalies disrupting the PPG waveform morphology are shown in Figures 1(b), (c) and (d). 
Accordingly, we posit that deeper analysis of PPG waveforms beyond just pulse rate can provide clues of cardiac function from a wearable device PPG sensor worn comfortably all the time. 
This has the potential to dramatically increase the clinical value of data from the current generation of PPG-based wearable devices. 
Firstly, cardiac abnormality clues identified in wearable PPG can be used to flag the patient for a thorough ECG investigation. Secondly, monitoring these types of events in broader populations may lead to new insights around their long-term and large-scale occurrence and impact in general populations.
Finally, these clues may provide a salient signal of cardiac function which can augment health deterioration early warning algorithms -- allowing them to make earlier and more accurate predictions for many serious health conditions.

%
%
%


%

%

\section{Related Work}
\label{sec:relatedwork}

%
Finding "surprising/interesting/unexpected/novel" sub-sequences in time series is generally referred to as \textit{anomaly detection} \citep{Lin:2003}. 
A wide body of literature originating in machine learning and statistics has developed generalisable and domain-specific anomaly detection and classification techniques, typically based on learning common patterns and expected statistical distributions \citep{Chandola:2009}.
%

Anomaly detection and classification approaches have long been applied to ECG and electroencephalography (EEG) signals.
For example, \cite{Rahhal:2016} used deep learning to identify cardiac events in ECG.
ECG data has very distinctive morphology (i.e., PQRST waveform complex), and many algorithms have been built into ECG monitors to automatically classify certain cardiac conditions such as ST elevation with high precision. In contrast, PPG contains less time and frequency domain information, and is more susceptible to calibration and motion artefact noise, so requires a different approach.
Specifically for recognising PPG waveforms impacted by artefacts, Signal Quality Indices (SQI) identify good and bad PPG signals \cite{Orphanidou:2015,Karlen:2012}. SQI are heuristic-based models, based on the timing of pulse waves and known physiological distributions. They classify larger fragments of signal as a binary good or bad, and are not designed to highlight specific anomalous regions of PPG signal.

\section{PPG Anomaly Detection}
\label{sec:adapproach}


\begin{figure}
   \centering
   \vspace{-0.3cm}
   \subfloat{\includegraphics[width=.42\textwidth]{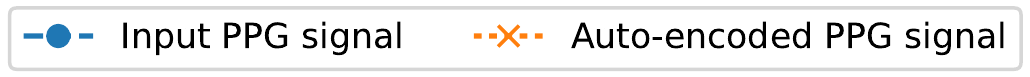}}\\
   \vspace{-0.3cm}
   \subfloat{\includegraphics[width=.44\textwidth]{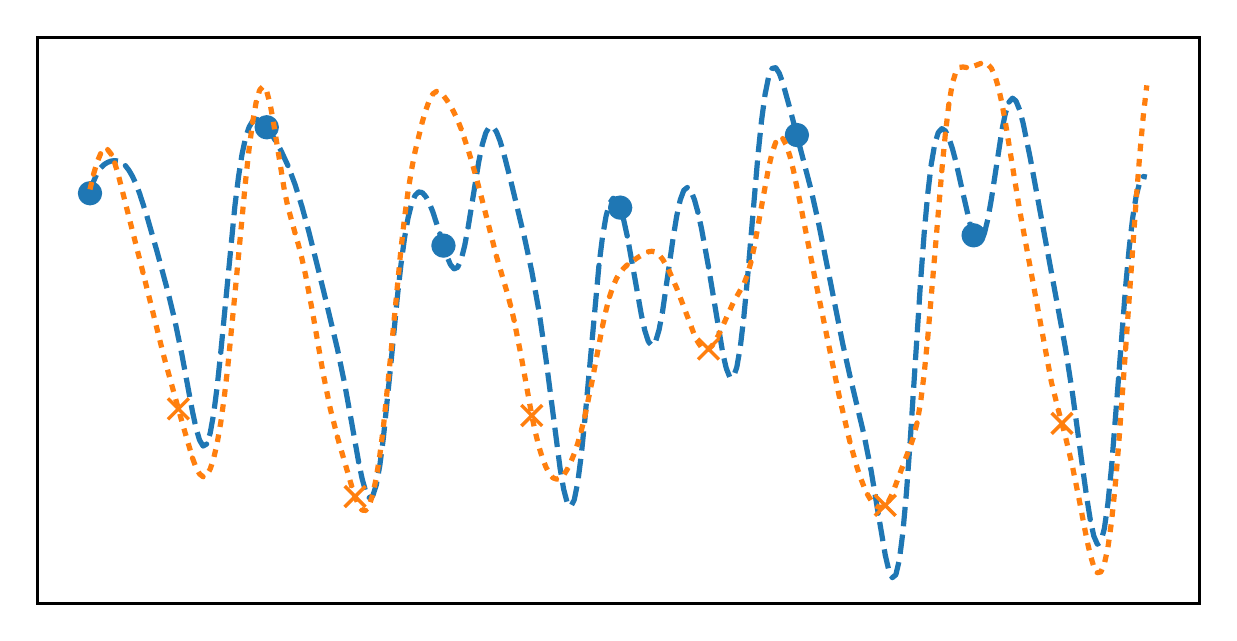}}\\
   \vspace{-0.5cm}
    \subfloat{\includegraphics[width=.44\textwidth]{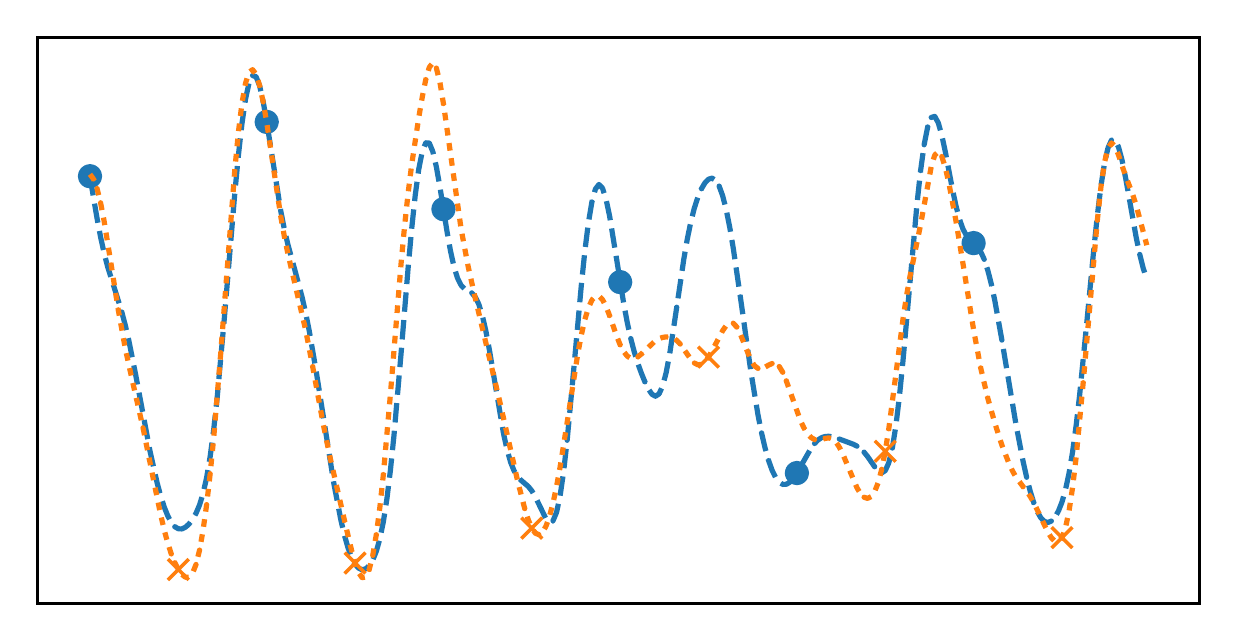}}
  \vspace{-0.2cm} 
   \caption{Input PPG and autoencoded counterpart examples. Note the autoencoder failing to reproduce anomalous regions.}
   \label{fig:autoencoderexamples}
\end{figure}

Artefacts in PPG can originate from many sources, including (i) physiological abnormalities (e.g., cardiovascular issues), (ii) motion corruption, and (iii) poor PPG calibration (i.e., how much light to transmit into the skin - which is affected by skin colour, circulation, adipose tissue and sensor contact pressure).
As we are interested in recognising physiological artefacts - our approach must first rule out that the source of any anomaly recognised is not from motion or calibration issues. 
Consequently, the snap40 upper-arm wearable device employs various continuous proprietary calibration routines which quickly recalibrate the PPG signal if it is compromised, e.g., the patient moves the device for comfort to a location on their arm with different tissue properties.
Moreover, the snap40 device discards PPG signal data when wearer motion will irretrievably corrupt the PPG signal. Since cardiac abnormality analysis is very sensitive to small perturbations in the waveform caused by motion, we set the motion filtering level to 'very still', in laying or sitting postures.

Our proposed cardiac abnormality detection approach comprises two stages. 
Firstly, we train an autoencoder to learn what typical normal PPG morphology and rhythm looks like.
To this end, we employ a deep recurrent neural network LSTM autoencoder which is capable of learning the time- and freqency-domain patterns found in clean PPG signal.
Secondly, we use this autoencoder to encode and then reconstruct the input PPG signal based on its reduced dimensionality representation. By measuring the differences between autoencoded and original PPG signals, we identify regions of the PPG signal that the constrained autoencoder representation fails to sufficiently reproduce -- i.e., the specific anomalies. This unsupervised approach allows us to recognise anomalies without needing to explicitly label an anomaly training set at scale.
%



\subsection{Deep-learnt Autoencoder Implementation}
\label{sec:autoencoder}

Abstractly, an autoencoder learns, without supervision, an optimal reduced dimensionality representation of its training examples given permitted representational complexity (i.e., the number of neurons and hidden layers). 
%
%
%
%
In this application, an autoencoder provides an unsupervised method of learning the common and defining morphological and rhythm patterns of PPG signals. Accordingly, anything that is atypical in the input PPG signal is not encoded/decoded in the subsequent autoencoder output.

Two examples of PPG signal including cardiac abnormalities, and their autoencoded counterparts, are presented in Figure \ref{fig:autoencoderexamples}. Note how the autoencoder adequately reproduces components which have time and frequency typically seen in regular PPG signal, however falters around unusual patterns -- it is this faltering that allows to us to automatically recognise and flag specific abnormal PPG regions.


So the autoencoder does not learn to represent and reproduce the anomaly patterns we wish to expose, it is essential it is trained using only clean PPG signal samples without abnormalities.
Accordingly, we constructed a dataset of approximately 400,000 8-second clean and regular PPG signal samples from 300+ real patients, bootstrapped by semi-supervised FFT-based frequency analysis on a random sample of around 2,000 PPG fragments. We do not explicitly label any specific anomalies.
As human physiology often follows many heavy-tailed distributions, and a deep-learnt model requires a lot of training examples to be effective, a training set of this scale is necessary to reliably include clinically expected pulse rates of 35-180 beats/min, and various common physiologies such as high and low, but not abnormal, heart rate variabilities.

Digital signal processing (DSP) filtering was used to clean, down-sample and normalise PPG signal for the autoencoder. Since PPG is a time series, the autoencoder is a sequence-to-sequence model, implemented as a long-short term memory (LSTM) recurrent neural network in Tensorflow \cite{Abadi:2016}. 
Preliminarily, we used a 2-layer (with 80/40 neurons) LSTM. However, optimal neural network architecture and training is entirely dependent on the device sensor characteristics, DSP pipeline used, run-time constraints, training computation availability, training data scale and desired sensitivity goals.


\subsection{Flagging Anomalous PPG Regions}
\label{sec:anomalyhighlighting}

\begin{figure}
   \centering
   \subfloat[][]{\includegraphics[width=.22\textwidth]{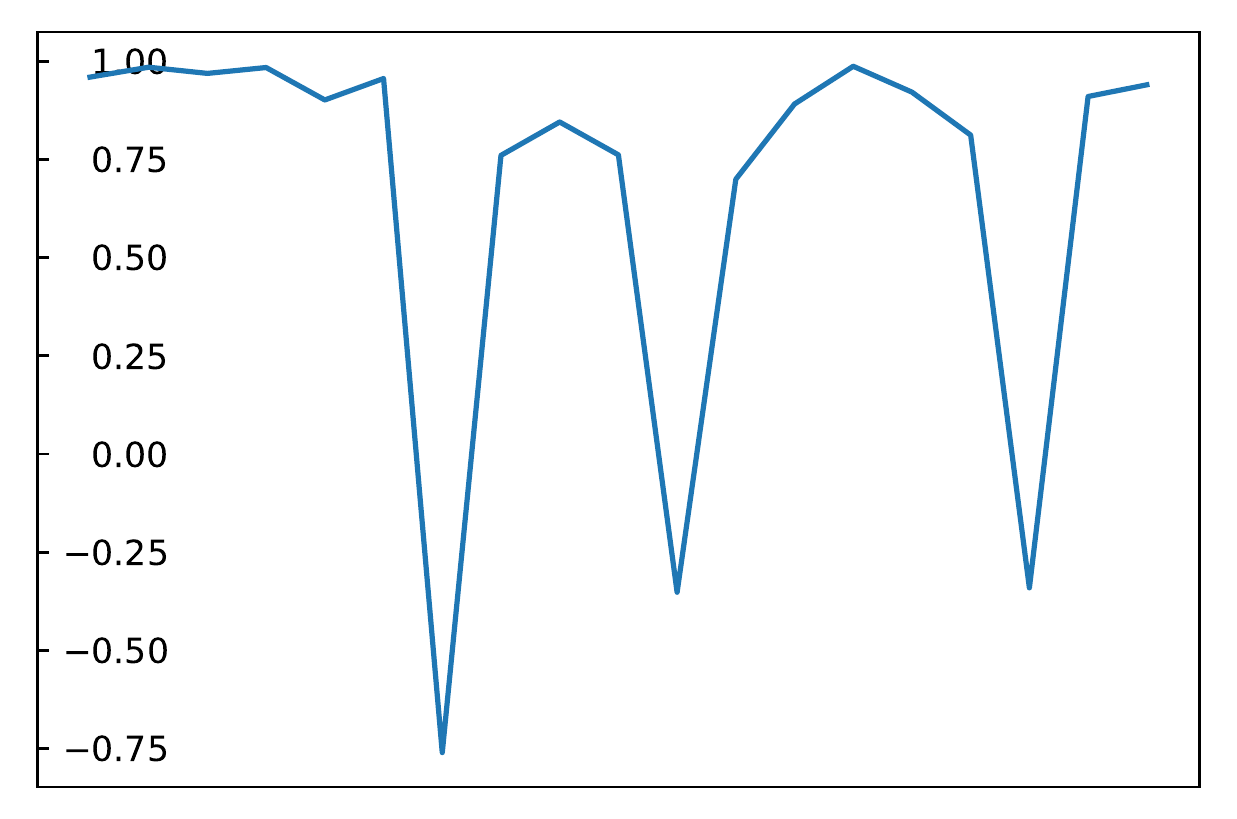}}\quad
    \subfloat[][]{\includegraphics[width=.22\textwidth]{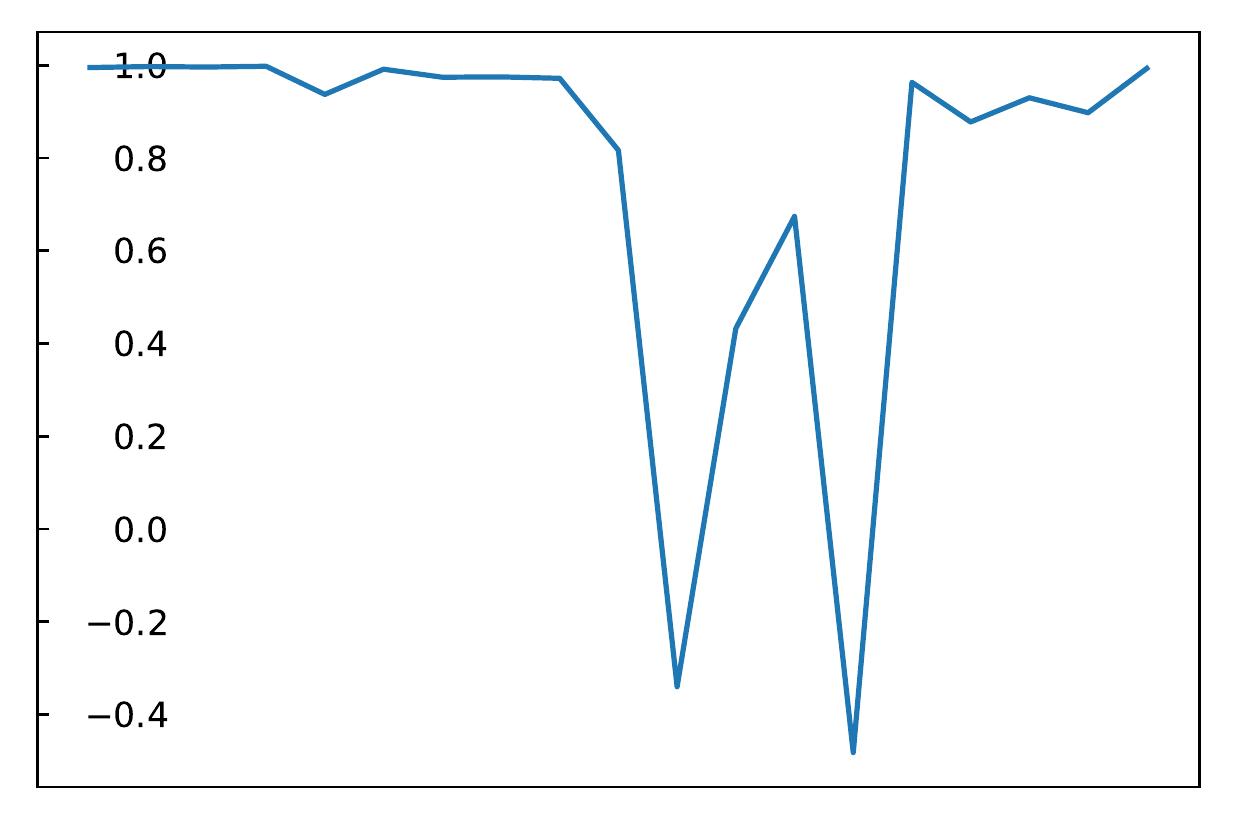}}\\
      \vspace{-0.2cm} 
       \subfloat[][]{\includegraphics[width=.22\textwidth]{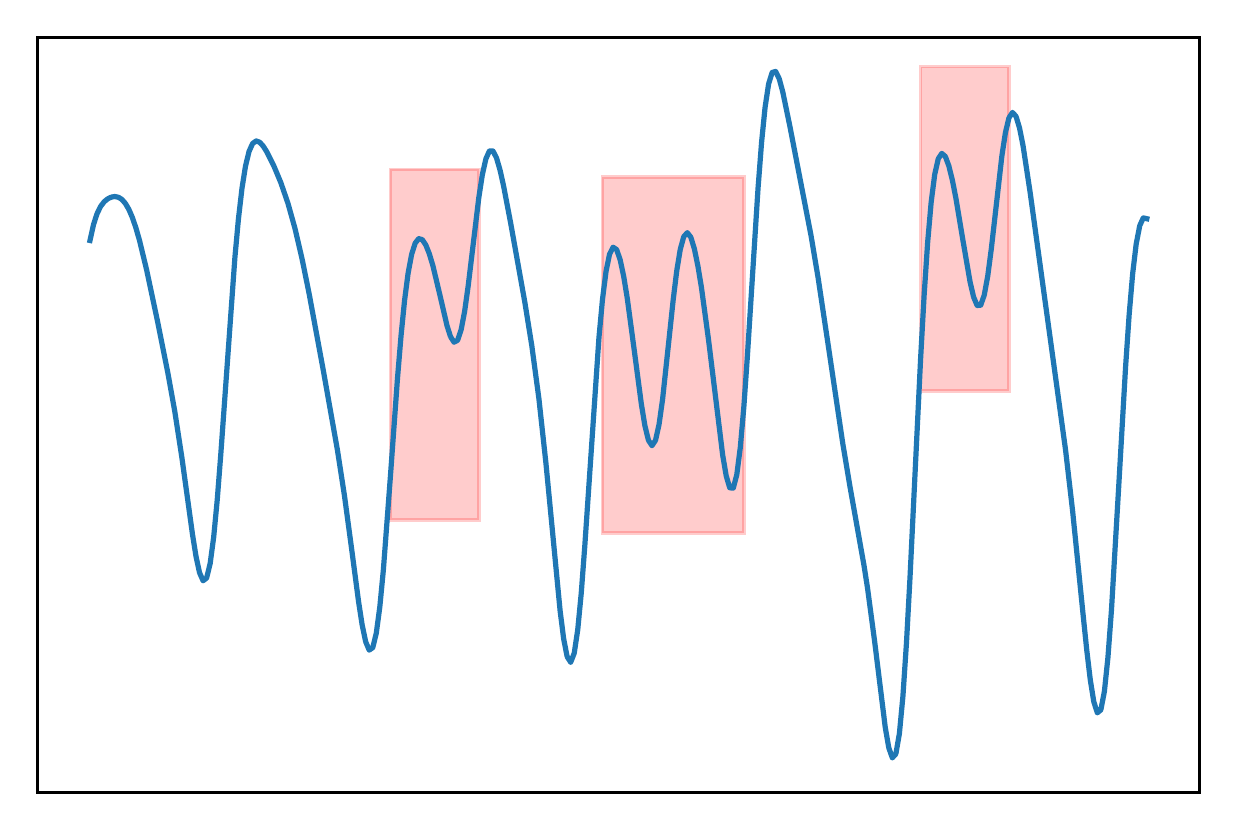}}\quad
    \subfloat[][]{\includegraphics[width=.22\textwidth]{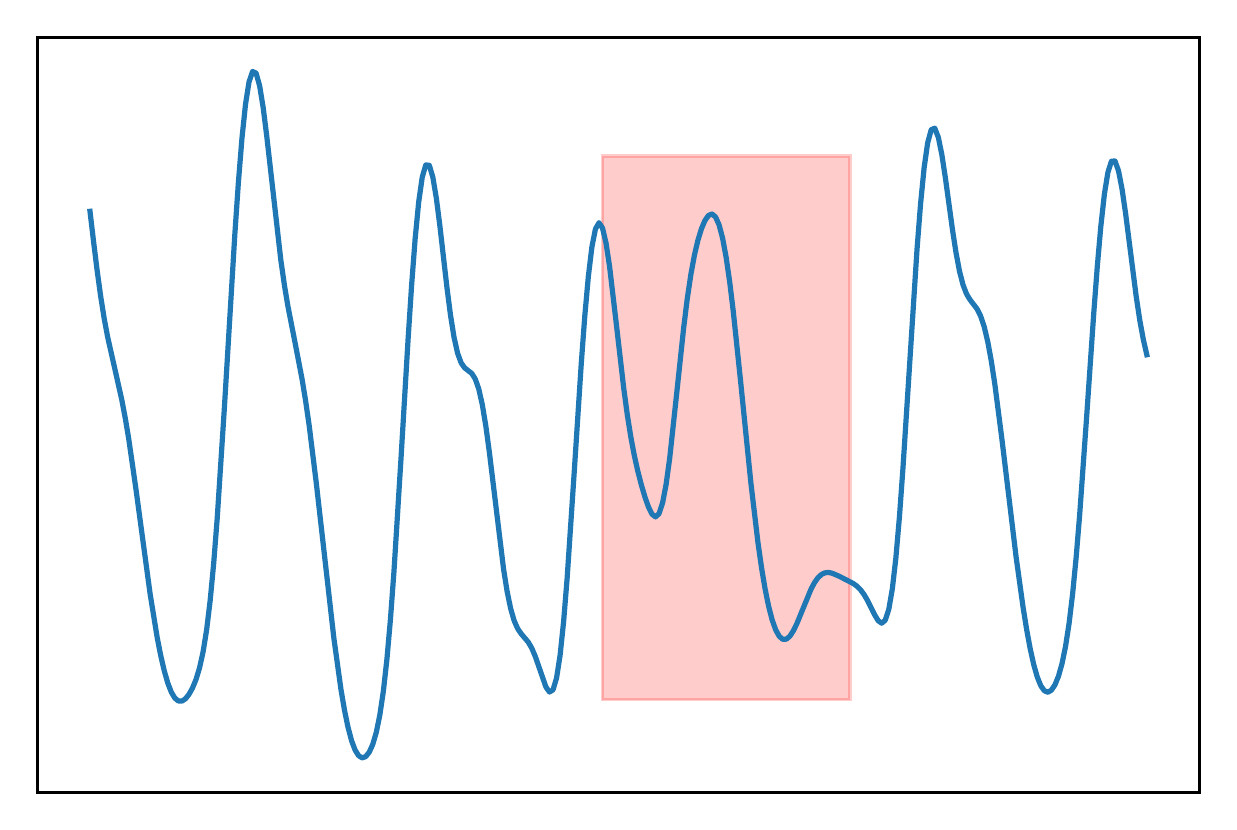}}
  \vspace{-0.2cm} 
   \caption{(a) and (b) show windowed Pearson's \textit{r} autoencoder reconstruction error from the PPG examples presented in Figure \ref{fig:autoencoderexamples}. Flagged PPG anomalies are shown in (c) and (d).}
   \label{fig:anomalousppgexamples}
\vspace{-0.48cm}
\end{figure}

Anomalous PPG signal regions are identified when the original PPG signal and its autoencoded counterpart diverge, as measured by a difference metric. Importantly for healthcare, this approach supports explainability as it explicitly flags specific irregular PPG signal regions for further automated analysis or manual human review.

We found absolute error as a difference metric is overly sensitive to occasional autoencoder underfitting for low frequency modulation in the PPG signal due to respiratory and parasympathetic induced variation \cite{Karlen:2013}. 
Future work to increase training data, using longer PPG length and neural network tuning will likely alleviate this.
Instead, computing Pearson's \textit{r} correlation co-efficient over sliding windows of the input PPG signal and its autoencoded counterpart was sufficient to reliably recognise anomalous regions.
Sensitivity of the anomaly detection is governed by both the size of these windows and the $r$ selected as an anomaly threshold. We selected half-second windows, with anomaly level $r < 0.6$ for experiments in Section \ref{sec:resultsdiscussion}.
%

In Figures \ref{fig:anomalousppgexamples}(a) and (b), we present the aligned autoencoder error, as measured by Pearson's \textit{r}. The respective Figures \ref{fig:anomalousppgexamples}(c) and (d), show the identified anomalies, defined as periods where \textit{r} falls below 0.6. Figure \ref{fig:anomalousppgexamples}(c) shows multiple similar anomalies manifesting rapidly, with two dominant faster/slower frequencies manifesting in the PPG signal. Meanwhile, Figure \ref{fig:anomalousppgexamples}(d) shows a single anomaly which is prominent compared to its surrounding PPG context.






\section{Experiment, Results \& Discussion}
\label{sec:resultsdiscussion}

To preliminarily evaluate and analyse the potential of our proposed approach, we investigate clinical and wearable sensor data from hundreds of clinical study patients with a range of pathologies and acuities being cared for in different settings in the UK and USA (e.g., in hospital wards, during surgery and at home).

Following recruitment,
patients wore the snap40 device on the upper-arm for one hour to ten days while undergoing standard care. The snap40 device passively captured low-motion green PPG sensor data, and wirelessly transmitted high fidelity waveforms for post-hoc analysis. 
The clinical studies and their respective patient populations were as follows: 
\textbf{HDU}: (medical/surgical high-dependency unit), comprises 120 patients who are high acuity and continuously monitored, wearing the device during the day. 
\textbf{AMU}: (acute medical unit) consists 250 patients higher acuity patients who have been recently admitted to hospital, wearing the device for up to 10 days, 24 hours a day. 
\textbf{SURGERY}: consists 30 peri- and post-operative general surgery patients, wearing the device for up to 2 days, 24 hours a day. 
\textbf{ED}: consists of 250 emergency department patients of varying acuity, wearing the device between 30-240 minutes, during the day. 
\textbf{HOME}: consists of 8 heart bypass patients for 2 days post-discharge to home care, wearing the device 24 hours a day.

In Section \ref{sec:ecgpvccomparison}, we first examine the accuracy and sensitivity of the proposed approach for recognising known cardiac abnormalities in PPG signal, compared to abnormalities (i.e., PVCs) recognised by a conventional bedside ECG monitor. Of course PVCs are only one type of cardiac abnormality, so this initial methodology provides an initial insight into performance using ECG-based gold standard (GS) PVC detection as a first proxy for real abnormal cardiac events. 
Since few patients receive continuous ECG monitoring, even in the hospital, this evaluation is based on the subset of HDU patients who had leads correctly attached for continuous ECG monitoring for whom we have data available. For these patients, we have a once-a-minute count of the zero or more PVCs detected in their ECG over that minute. 
We align available wearable PPG sensor data over each of those minutes, and compute set-based comparative evaluation measures (i.e., classification accuracy) based on the presence or absence of anomalies detected in the aligned PPG, and respective GS PVCs. 
We filter GS PVCs to those where there is at least 30-seconds of high quality aligned wearable PPG signal available. 

Following this, in Section \ref{sec:patientpopulationanalysis} to understand how cardiac abnormalities manifest at large in PPG signal collected from diverse patient populations, we analyse the overall and per-patient frequency of the cardiac abnormalities detected with our proposed approach.

\subsection{Comparative ECG Gold Standard Evaluation}
\label{sec:ecgpvccomparison}

Our test dataset contains 29 patients with a total of 2,852 zero or more PVCs/min GS observations (i.e., 47.6 hours ECG monitoring, averaging 98 minutes of observations per patient), where each GS observation has $\geq 30$ seconds of aligned good quality PPG signal. 

2,465 (86.4\%) of the GS PVCs/min are 0 PVCs/min observations, 387 (13.6\%) are >= 1 PVCs/min observations 
while 195 (6.8\%) are >= 2 PVCs/min observations.
At the extreme, one patient has a GS observation with 7 PVCs/min; showing PVCs are heavily skewed.

In Table \ref{tab:confusionmatrix-gte1pvcs}, we present the set-based detection accuracy confusion matrix for recognising GS PVCs (when there is >= 1 PVC/min) in PPG signal using our proposed approach. 
Our proposed approach successfully recognises around 60\% of PVCs in the PPG signal alone, and incorrectly recognises 23\% of PPG signals without a PVC as having a cardiac abnormality. This may be a genuine approach error, or it could be another type of cardiac abnormality.
%
Additionally, because of our high-motion PPG filtering, we do not have complete PPG signal coverage over the gold standard periods, so may miss some PVCs with this methodology limitation.
Furthermore, the gold standard itself will also have PVC classification error. 
%
\cite{Kurka:2015} found of 22,509 arrhythmia alarms analysed, 27.4\% where false alarms which grew to 91.4\% for acute life-threatening alarms, with no events missed -- indicating a preference for false positives rather than false negatives, thus affecting our evaluation metrics.

Allowing 1 or more PVCs/min in the GS means cardiac abnormalities are rare for the majority of GS observations. Accordingly, we compute the confusion matrix when there are 2 or more PVCs/min present in the GS. This increases true positives to 132 (68\%) 
- showing that when cardiac abnormalities are more prevalent, our PPG-based approach is increasingly more effective in recognising them.


Overall these initial results are very encouraging as they demonstrate that even with a limited evaluation methodology, a basic model can achieve reasonable sensitivity while maintaining specificity. 
Future work will investigate more robust methods to identify anomalies in autoencoder output, and if possible, classify the specific cardiac events which caused them.











\subsection{Patient Population Analysis}
\label{sec:patientpopulationanalysis}

We use our proposed approach to identify cardiac abnormalities in over ten thousand hours of PPG data randomly sampled from several large-scale clinical studies using the snap40 wearable device. Analysis results are presented in Table \ref{tab:studycardiacabnormalities}.


Patient demographics and biases can explain many of the cardiac abnormality differences between populations.
Expectedly, the often older and higher acuity patients in HDU and AMU had the most PPG samples with cardiac anomalies (i.e., 5.1\% and 5.8\%, respectively). 
In contrast, recently discharged patients at HOME had fewer PPG samples with cardiac abnormalities overall (i.e., 3.52\%) - and with lesser variability indicating more stable cardiac health compared to in-patients. Likewise, ED has a wide range of patient acuities (indeed many will go to AMU), so while it has on average a lower PPG-based cardiac abnormality occurrence (i.e., 3.1\%), some patients have far more as shown by the large variability (i.e., $\pm7.9\%$).
Interestingly, SURGERY patients had the fewest PPG samples with cardiac abnormalities, perhaps due to selection of fitter patients eligible for surgery, close monitoring or perhaps therapeutic effects.

\begin{table}[ht]
\small

\begin{center}
\begin{tabular}{ |c|c|c| } 
 \hline
   & Anomaly in PPG \cmark & Anomaly in PPG \xmark \\ 
 \hline
 PVC in ECG \cmark & \makecell{231 \\(59.7\% \textit{true positive})} & \makecell{156 \\(40.0\% \textit{false negative})} \\ 
 \hline
 PVC in ECG \xmark & \makecell{574 \\(23.2\% \textit{false positive})} & \makecell{1,891 \\(76.7\% \textit{true negative})} \\ 
 \hline
\end{tabular}
\end{center}






\caption{Set-based detection accuracy confusion matrix for recognising ECG-based GS PVCs (when there is >= 1 PVC/min) in aligned PPG signal using our proposed approach.}
\label{tab:confusionmatrix-gte1pvcs}
\vspace{-0.6cm}
\end{table}
\vspace{-0.4cm}
\begin{table}[ht]
\small

\begin{center}
\begin{tabular}{ |c|c|c| } 
 \hline
& \multicolumn{2}{ |c| }{ \makecell{ \% of PPG samples with anomalies, \\per patient}}  \\ 
 \hline
   Population: & Avg (Stdev) & Max  \\ 
 \hline
 HDU & 5.10\% ($\pm5.0\%$) & 17.4\% \\ 
 \hline
 SURGERY & 1.54\% ($\pm2.2\%$) & 6.9\% \\ 
 \hline
 ED & 3.17\% ($\pm7.9\%$) & 59.1\% \\ 
  \hline
 AMU & 5.75\% ($\pm11.9\%$) & 61.5\% \\  
  \hline
 HOME & 3.52 ($\pm1.9\%$) \% & 5.8\% \\ 
 \hline
\end{tabular}
\end{center}
 \caption{Frequency of cardiac abnormalities recognised in PPG samples, aggregated per patient, in each patient population.}
 \label{tab:studycardiacabnormalities}
 \vspace{-0.4cm}
\end{table}
%
%
%
%
%
\section{Conclusion}
\label{sec:conclusion}

Cardiac abnormalities such as ectopic beats and AF manifest in a wearable device PPG waveform as they disrupt blood flow. Some of these abnormalities are benign, while others can be a serious health risk factor. Accordingly, identifying cardiac abnormalities in PPG signals provided by a conveniently practical wearable device, as opposed to conventional inconvenient ECG monitors requiring multiple electrodes and leads can be valuable.

We demonstrated that cardiac abnormalities, where PPG signal deviates from typical morphology and rhythm, can be recognised in wearable PPG signal with an unsupervised deep-learnt autoencoder anomaly detection approach.
Preliminary evaluation on a large ECG-based gold standard dataset showed our approach recognises 60\%+ of ECG-detected PVCs in PPG signal, with a false positive rate of 23\%. Expectedly, analysis of several large clinical study datasets showed cardiac abnormalities detected are more frequent in higher acuity patients.
Future work will enhance accuracy and sensitivity, and explore specific cardiac event classification.


\bibliographystyle{abbrv}
\small
\bibliography{bibliography}

\begin{thebibliography}{1}

\bibitem{Abadi:2016}
M.~Abadi, et. al. 
\newblock Tensorflow: A system for large-scale machine learning.
\newblock In OSDI'16, pages 265--283, Berkeley, CA,
  USA, 2016. USENIX Association.

\bibitem{Chandola:2009}
V.~Chandola, A.~Banerjee, and V.~Kumar.
\newblock Anomaly detection: A survey.
\newblock {\em ACM Comput. Surv.}, 41(3):15:1--15:58, July 2009.

\bibitem{Karlen:2012}
W.~Karlen, K.~Kobayashi, J.~M. Ansermino, and G.~A. Dumont.
\newblock Photoplethysmogram signal quality estimation using repeated gaussian
  filters and cross-correlation.
\newblock {\em Physiological Measurement}, 33(10):1617, 2012.

\bibitem{Karlen:2013}
W.~Karlen, S.~Raman, J.~M. Ansermino, and G.~A. Dumont.
\newblock Multiparameter respiratory rate estimation from the
  photoplethysmogram.
\newblock {\em IEEE TBME}, 60:1946--1953,
  2013.

\bibitem{Kurka:2015}
N.~Kurka, T.~Bobinger, B.~Kallm{\"u}nzer, J.~Koehn, P.~D. Schellinger,
  S.~Schwab, and M.~K{\"o}hrmann.
\newblock Reliability and limitations of automated arrhythmia detection in
  telemetric monitoring after stroke.
\newblock {\em Stroke}, 46(2):560--563, 2015.

\bibitem{Lin:2003}
J.~Lin, E.~Keogh, S.~Lonardi, and B.~Chiu.
\newblock A symbolic representation of time series, with implications for
  streaming algorithms.
\newblock In {\em ACM SIGMOD 2003}, DMKD '03, pages 2--11, New York, NY,
  USA, 2003. ACM.

\bibitem{Orphanidou:2015}
C.~Orphanidou, T.~Bonnici, P.~Charlton, D.~Clifton, D.~Vallance, and
  L.~Tarassenko.
\newblock Signal-quality indices for the electrocardiogram and
  photoplethysmogram: Derivation and applications to wireless monitoring.
\newblock {\em IEEE JBHI},
  19(3):832--838, 2015.

\bibitem{Rahhal:2016}
M.~A. Rahhal, Y.~Bazi, H.~AlHichri, N.~Alajlan, F.~Melgani, and R.~Yager.
\newblock Deep learning approach for active classification of electrocardiogram
  signals.
\newblock {\em Information Sciences}, 345:340 -- 354, 2016.

\end{thebibliography}

\end{document}